\documentclass[floatfix,twocolumn,secnumarabic,amssymb, amsmath,nobibnotes,aps,prb,superscriptaddress]{revtex4-1}

\usepackage{graphicx}
\usepackage{dcolumn}
\usepackage{bm}
\usepackage{amsthm,amsmath}
\usepackage{multirow}

\usepackage{tikz}
\usepackage{chemfig}
\usepackage[T1]{fontenc}
\usepackage{lmodern}
\tikzstyle{format}=[draw, thin, rounded corners=5pt, line width =0.5, fill=concepts!10!white]
\usetikzlibrary{patterns}

\begin{document}

\title{Assessing The Band Gap Problem By Improving Upon The Semilocal Exchange Hole Potential} 

\author{Subrata Jana}
\altaffiliation{Corresponding author: subrata.jana@niser.ac.in}
\affiliation{School of Physical Sciences, National Institute of Science Education and Research,
HBNI, Bhubaneswar 752050, India}
\author{Hemanadhan Myneni}
\altaffiliation{Corresponding author: hemanadhiitk@gmail.com}
\affiliation{Department of Physics and Astronomy, University of Delaware, Newark, Delaware 19716, United States}
\author{Prasanjit Samal}
\altaffiliation{Corresponding author: psamal@niser.ac.in}
\affiliation{School of Physical Sciences, National Institute of Science Education and Research,
HBNI, Bhubaneswar 752050, India}

\date{\today}
             
\begin{abstract}
An asymptotic corrected exchange hole potential analogous to Becke-Roussel [\textcolor{blue} {A. D. Becke 
and M. R. Roussel, Phys. Rev. A {\bf{39}}, 3761 (1989)}] is constructed by modeling the exchange hole 
using the generalized coordinate transformation based on density matrix expansion. The model potential 
is Laplacian free and the inhomogeneity present in the system is included in the momentum vector 
without affecting the uniform density limit. The parameters associated with the model exchange hole are 
fitted with the spherical atoms. The newly constructed potential along with Tran-Blaha modified 
Becke-Johnson (TBMBJ) [\textcolor{blue} {F. Tran and P. Blaha, Phys. Rev. Lett. 102, 226401 (2009)}] potential 
quite accurately produces the band gap of various materials ranging from semiconductor through insulators. 
The results for band gap are improved compared to the TBMBJ and other standard semilocal exchange functionals.   
\end{abstract}
\maketitle

\section{Introduction}\label{intro}
The Kohn-Sham (KS) ground-state density functional theory (DFT)~\cite{hk64,ks65} has become the {\it{de facto}} 
standard for the electronic structure calculations in physics, chemistry and materials science. The tremendous 
success of the KS-DFT lies in the accurate approximations to the exchange-correlation (XC) energy functionals
(E$_{\textit{XC}}$) or the corresponding potentials (V$_{\textit{XC}}$) that are being developed over decades.
The systematic improvement of the exchange-correlation functional is still a very active research field. The 
hierarchy of density functional approximations were often represented by rungs of a Jacob's Ladder, where each 
rung of the ladder introduces an additional ingredient to the energy density. On the lowest rung of the ladder 
is the local density approximation (LDA) which depends only on the electron $\rho(\mathbf{r})$. On the next 
rung of the ladder is the generalized gradient approximations~\cite{b83,jp85,pw86,b88,lyp88,b3pw91,VLB,pbe96,
pby96,ae05,pbesol,lc1}, where the electron density $\rho(\mathbf{r})$ and its gradient, $\mathbf{\nabla}
\rho(\mathbf{r})$ are the basic inputs. As a successor to it, the next level of sophistication comes through 
the "meta generalized gradient approximations" (meta-GGAs),~\cite{vsxc98,hcth,tpss,mO6l,akk,revtpss,lc2,lc3,
scan15,tm16} which uses the positive KS kinetic energy density $\tau(\mathbf{r})$ = $1/2\sum_i{|{\bf{\nabla}}
\psi_i(\mathbf{r})|}^2$ (with $\psi_i$ being the occupied KS orbitals). Meta-GGAs are the most appealing and 
accurate semilocal functionals for solids due to satisfaction of some accurate constraint such as slowly 
varying density correction. The meta-GGAs are implemented for solids by replacing the Laplacian of density by 
the slowly varying density approximation of the KS kinetic energy density~\cite{tpss,scan15,tm16}. Recent study 
shows that replacement of Laplacian by approximation of kinetic energy density at least qualitatively very good 
to some degree \cite{kejcp}. Several meta-GGA functionals with increasing accuracy are proposed during recent 
years for electronic structure calculation for solids, materials and molecules~\cite{tpss,scan15,tm16}. While 
these functionals are sufficiently accurate to describe ground state properties but failed in the certain limit, 
especially in describing the excited state properties such as charge-transfer, ionization potential, electron 
affinity, band gap, response properties and electron transport in solids. Standard semilocal functionals failed 
in this regard. Several attempts, therefore, made to improve the band gap such as self interaction correction (SIC)
~\cite{JP1979,PC1988,PZ1981}, Hubbard correction over density functional formalism (DFT + U)~\cite{AZA1991}, DFT 
based dynamic mean field theory (DFT+DMFT), mixing non-local Hartree-Fock with semilocal functional and make 
them hybrid~\cite{pbe0,jcphse,PMHKGA2006,GAMK2007,ASJBNK2013,AJSNK2015}, quasi particle green function correction 
(GW)~\cite{ASG2000,FSK2004,SKF2006,CSK2007,SK2007,SMK2007}. All methods except DFT+U are computationally expensive 
and DFT+U is only applicable to localized electrons such as $3d$ or $4f$. 

The failure of band gap prediction within the standard KS formalism using semilocal XC energy functional can be 
understood as '' the failure of describing the derivative discontinuity ''~\cite{PPLB1982,SS1983,PL1983,DJT2003,
CJCS1998,KK2014} and '' the delocalization error''~\cite{dloc1,CC2010}. The derivative discontinuity can be 
understood as follows: The band gap in KS formalism is the difference between the orbital energy of the highest
occupied (HO) orbital i.e., conduction band (CB) to the lowest unoccupied (LU) orbital i.e., valence band (VB) 
($\Delta_g^{KS}=\varepsilon_{HO}-\varepsilon_{LU}=\varepsilon_{CB}-\varepsilon_{VB}$). The fundamental band gap 
is defined as the difference between ionization potential ($IP$) and electron affinity ($EA$) i.e, $E_g=-\{[E_0^N
-E_0^{N-1}]-[E_0^{N+1}-E_0^N]\}=IP-EA$. Using the variational formalism it can be shown that there exists an extra
derivative discontinuity in XC functional ($\Delta_{xc}$), which makes band gap in KS formalism different from 
the fundamental band gap as $E_g=\Delta_g^{KS}+\Delta_{xc}$. Therefore, error in $\Delta_{xc}$ results $\Delta
_g^{KS}$ differ from $E_g$. It is noteworthy to mention that semilocal XC functionals do not show any derivative 
discontinuity and therefore, perform poorly in band gap prediction. Several resolutions have been proposed 
including exact exchange formalism which by construct posses $\Delta_{xc}$ \cite{GKG1997,ED2011,TK1995,gor1,gor2,
gor3,gk1}. However, due to the high computational cost of exact exchange for the extended system, it has been proposed 
to mimic the exact-exchange (EXX) by semilocal GGA or meta-GGA type exchange potential~\cite{BJ2006,TBS2007,tbmbj,
KOER2010,AK2013}. A simple effective potential which is the promising substitution of EXX is the Becke-Johnson (BJ) 
meta-GGA potential~\cite{BJ2006}. The performance of this potential for atomic system shows that highest occupied 
eigenvalue corresponds to IP. Albeit promising substitution of EXX, the band gap prediction using BJ potential is
still not very accurate~\cite{TBS2007}. A simple modification of BJ potential using optimized screening parameter 
reduced the error between the large and small band gap solids. The main motivation behind this screening parameter 
comes from the range separated HSE hybrid functional \cite{hsescre}. Next, the Tran-Blaha's modification over 
BJ potential (TBMBJ)~\cite{tbmbj} potential, produces improved results for semiconductor band gaps. The 
first term of TBMBJ potential contains Becke-Roussel (BR) potential~\cite{br89}. Whereas, the second term contains
orbital shifting KS kinetic energy dependent term. 

The BR potential is the semilocal substitution of the Slater potential with correct asymptotic behavior, obtained 
from the Taylor series expansion of the exchange hole. Thus, in this everything rely on the modeling of the exchange
hole. One of the ways of obtaining the exchange hole is the density matrix expansion (DME)~\cite{tsp03,tm16}. Recently, 
DME based exchange energy functional construction has gained momentum due to its correct formal properties such as  
(i) correct uniform density limit of exchange hole, (ii) exactly obtaining Becke's expansion and (iii) satisfaction
of convergence criteria without any real space cutoff~\cite{tsp03,tm16}. Inspired by these attempts, we have proposed 
here a slightly different form of Becke's exchange hole expansion, which in combination with the BJ and TBMBJ produces 
much-improved results for band gap. Our proposed exchange hole based on generalized gradient expansion is free from 
the Laplacian of density as is implemented in meta-GGA functionals~\cite{tpss,tm16}. Also, the inhomogeneity present
in the system is included in Fermi momentum vector without affecting the uniform density limit of exchange hole. 
Our paper is organized as follows. In the following section, we will briefly discuss the generalized coordinate 
transformed exchange hole model. Then, the section next to it, based on our proposed exchange hole we will propose 
the improved BR potential. Subsequently, we will fix the parameters present in the proposed functional by making a 
comprehensive study of atomic systems. Finally, to test the performance of the functional and comparison of results
obtained, we will calculate the band gap of semiconductors along with other semilocal and nonlocal functionals.

\section{Theoretical Background}
\label{seci}
The exchange energy can be regarded as the electrostatic interaction between the electron density $\rho_{\sigma}
(\mathbf{r})$ at reference position $\mathbf{r}$ with spherical averaged exchange hole density $\langle\rho_
{x\sigma}(\mathbf{r},\mathbf{r} + \mathbf{u})\rangle$ at $\mathbf{r}+\mathbf{u}$, where $\mathbf{u}$ is the 
separation between two electrons. So
\begin{equation}
E_{x} = -\frac{1}{2}\sum_{\sigma}\int\int\frac{\rho_{\sigma}(\mathbf{r})\langle\rho_{x\sigma}(\mathbf{r},
\mathbf{r} + \mathbf{u})\rangle}{u}~d\mathbf{r}d\mathbf{u}.
\label{sec1eq1}
\end{equation}
The exchange hole density depends not only on the electronic separation ($\mathbf{u}$) but also on the orientation 
of it. The spherically averaged exchange hole density in Eq.(\ref{sec1eq1}), can be expressed in terms of the $1^{st}$ 
order Hartree-Fock reduced density matrix as
\begin{equation}
\langle\rho_{x\sigma}(\mathbf{r},\mathbf{r}+\mathbf{u})\rangle = -\frac{\langle|\varGamma_{1\sigma}(\mathbf{r},
\mathbf{r} + \mathbf{u})|^{2}\rangle}{\rho_{\sigma}(\mathbf{r})}~.
\label{sec1eq2}
\end{equation}
On the other hand, the spherically averaged $1^{st}$ order reduced density matrix is related to reduced density matrix 
as
\begin{equation}
\langle|\varGamma_{1\sigma}(\mathbf{r},\mathbf{r}+\mathbf{u})|\rangle = \frac{1}{4\pi}\int \varGamma_{1\sigma}
(\mathbf{r},\mathbf{r} + \mathbf{u})~d\Omega_u~.
\label{sec1eq3}
\end{equation}
Also, the reduced density matrix is expressed in terms of KS orbitals as
\begin{equation}
\varGamma_{1\sigma}(\mathbf{r},\mathbf{r}+\mathbf{u}) = \sum_{i}^{\sigma occ}\psi_{i\sigma}^*(\mathbf{r})
\psi_{i\sigma}(\mathbf{r} + \mathbf{u})~.
\label{sec1eq5}
\end{equation}
The exchange hole potential generated by the spherical averaged exchange hole at reference point $\mathbf{r}$ is,
\begin{equation}
U_{x\sigma}(\mathbf{r})=-\int\frac{\langle\rho_{x\sigma}(\mathbf{r},\mathbf{r}+\mathbf{u})\rangle}{u}~d\mathbf{u},
\label{sec1eq6}
\end{equation}
Thus exchange hole is an important concept as knowing the exchange hole, exchange potential can be modeled.
Following the expression of exchange potential, the exchange energy expression i.e. Eq.(\ref{sec1eq1}) 
is defined as
\begin{equation}
E_x = \frac{1}{2}\sum_{\sigma}\int\rho_{\sigma}(\mathbf{r})U_{x\sigma}(\mathbf{r})d\mathbf{r}~.
\label{sec1eq7}
\end{equation}
Thus, knowing exchange potential one can obtain exchange energy. Here, it is important to note that the associated
exchange potential is not the functional derivative of exchange energy. But, the following two important properties 
(i) it obey sum rule and (ii) negativity criteria i.e.
\begin{eqnarray}
\int\langle\rho_{x\sigma}(\mathbf{r},\mathbf{r}+\mathbf{u})\rangle d\mathbf{u}&=&-1\nonumber\\
\langle\rho_{x\sigma}(\mathbf{r},\mathbf{r}+\mathbf{u})\rangle&\leq& 0
\label{sec2eq8}
\end{eqnarray}
are satisfied by the exchange hole.

\section{Construction of Exchange Hole, Potential and Energy}
Based on the generalized coordinate transformed DME proposed by Tao et. al.~\cite{tsp03}, the $1^{st}$ order reduced 
density matrix is defined as
\begin{equation}
\varGamma_{1\sigma}^{t}(\mathbf{r},\mathbf{r}+\mathbf{u}) = \sum_i^{\sigma occ}\psi_{i\sigma}^*(\mathbf{r} + 
(\lambda-1)\mathbf{u})\psi_{i\sigma}(\mathbf{r}+\lambda\mathbf{u}).
\label{sec1eq9}
\end{equation}
The corresponding spherically averaged exchange hole,
\begin{equation}
\langle\rho^{t}_{x\sigma}(\mathbf{r},\mathbf{r}+\mathbf{u})\rangle = -\frac{\langle|\varGamma^{t}_{1\sigma}
(\mathbf{r},\mathbf{r} + \mathbf{u})|^{2}\rangle}{\rho_{\sigma}(\mathbf{r})},
\label{sec1eq10}
\end{equation}
where $\lambda$ is the coordinate transformed parameter that takes values from $1/2$ to $1$. The $\lambda=1/2$ 
corresponds to the maximally localized exchange hole. Whereas, $\lambda = 1$ gives the conventional exchange hole 
model~\cite{tsp03}. Now the coordinate transformed exchange hole (Eq.\ref{sec1eq10}) can be used to obtain exchange 
potential and exchange energy
\begin{equation}
U^t_{x\sigma}(\mathbf{r})=-\int\frac{\langle\rho^t_{x\sigma}(\mathbf{r},\mathbf{r}+\mathbf{u})\rangle}{u}~
d\mathbf{u}
\label{sec1eq11}
\end{equation} 
and
\begin{equation}
E^t_x = \frac{1}{2}\sum_{\sigma}\int\rho_{\sigma}(\mathbf{r})U^t_{x\sigma}(\mathbf{r})d\mathbf{r}
\label{sec1eq12}
\end{equation}
respectively. So, we have proposed an exchange hole model: (i) based on generalized coordinate transformation and (ii) 
it includes inhomogeneity through Fermi momentum without hindering homogeneous limit of exchange hole. In the present
proposition, the small $\bf u$ expansion of the exchange hole becomes,
\begin{equation}
\begin{split}
\langle\rho^t_{x\sigma}(\mathbf{r},\mathbf{u})\rangle = \rho_{\sigma}(\mathbf{r})+ \frac{u^2}{6}
\Big[2(\lambda^2-\lambda+\frac{1}{2})\nabla^2\rho_{\sigma}(\mathbf{r})-4\tau_{\sigma}\\
+\frac{6}{5}k_{\sigma}^2
\rho_{\sigma}(\mathbf{r})(f_{\sigma}^2-1)+\frac{1}{2}(2\lambda-1)^2\frac{(\vec{\nabla}\rho_{\sigma}
(\mathbf{r}))^2}{\rho_{\sigma}(\mathbf{r})}\Big],
\end{split}
\label{hole}
\end{equation}
with,
\begin{equation}
\begin{split}
 f_{\sigma} = \Big[1+10(\frac{70}{27})\frac{1}{4(6\pi^2)^{\frac{2}{3}}}(2\lambda-1)^2x^2_{\sigma} + \\
\frac{\beta}{16(6\pi^2)^{\frac{4}{3}}}(2\lambda-1)^4x_{\sigma}^4\Big]^{\frac{1}{10}},
\end{split}
\end{equation}
where $x_{\sigma} = |\nabla\rho_\sigma|/\rho_\sigma^{\frac{4}{3}}$ is the reduced density gradient. The exchange
hole expression is given in Eq.(\ref{hole}) is obtained by small $u$ expansion of Tao-Mo \cite{tm16} DME exchange hole. 
The parameters $\lambda$ along with $\beta$ will be determined later. Here, the expansion of Becke is recovering by 
considering $f_\sigma\approx1$ and $\lambda=1$ i.e., slowly varying density with conventional exchange hole. Our 
present model also correctly recovers the uniform model exchange hole considering expansion up to $u^2$. The purpose 
of including inhomogeneity through $k$ is coming from DME of exchange hole~\cite{tm16}. The $k=k_F$, corresponds to 
the exchange hole of uniform exchange hole. But for inhomogeneous systems, one should vary $k$ from its homogeneous 
counterpart part by including the inhomogeneity parameter $f_\sigma$, which becomes $1$ for the homogeneous system. 
The TM-DME exchange hole~\cite{tm16} has fixed the form of $f_\sigma$ using normalization of exchange hole. Whereas,
present modification keeps intact the uniform density limit of BR potential because homogeneity of the system is 
transnationally invariant under generalized coordinate transformation. It is necessary to recover the homogeneous 
exchange potential as a limiting case in case of homogeneous system as mentioned by Tran-Blaha~\cite{tbmbj}. This is 
the main motive to use BR potential instead of Slater potential in TBMBJ. Also computationally, the BR potential is 
a couple of times faster than the Slater potential. In other words, asymptotic nature of the BR potential is useful 
to correctly capture the long range excitation effects. For slowly varying density, $k=k_F$ and $f_\sigma\approx 1$.
But for inhomogeneous systems, one needs to modify the Thomas-Fermi wave vector accordingly. Thus, inclusion of
inhomogeneity information through the Thomas-Fermi wave vector is essential and is done in our approach. Now, we 
have defined $Q_\sigma$ according to Becke's way as,
\begin{eqnarray}
Q_{\sigma} = \frac{1}{6}\Big[2(\lambda^2-\lambda+\frac{1}{2})\nabla^2\rho_{\sigma}(\mathbf{r})+\frac{6}{5}
k_{\sigma}^2 \rho_{\sigma}(\mathbf{r})(f_{\sigma}^2-1)\nonumber  \\
-2\gamma D_{\sigma}\Big]\nonumber\\
\end{eqnarray}
with $D_{\sigma}=2\tau_{\sigma}-\frac{1}{4}(2\lambda-1)^2\frac{(\vec{\nabla}\rho_{\sigma}(\mathbf{r}))^2}
{\rho_{\sigma}(\mathbf{r})}$. The adjustable parameter $\gamma$ was originally proposed by Becke-Roussel \cite{br89}. 
In the present case, $\gamma$ will be fixed later in this paper. The presence of Laplacian in the above expression
makes the exchange hole diverge, especially near nucleus. So, the Laplacian present in $Q_\sigma$ will be replaced 
using the semi-classical approximation of kinetic energy density i.e.,   
\begin{equation}
\nabla^2\rho_{\sigma}(\mathbf{r}) = 3[2\tau_{\sigma}-\tau_\sigma^{unif}-\frac{1}{36}\frac{(\vec{\nabla}
\rho_{\sigma}(\mathbf{r}))^2}{\rho_{\sigma}(\mathbf{r})}].
\end{equation}
Using this modification our present $Q_\sigma$ becomes,
\begin{eqnarray}
Q_{\sigma} = \frac{1}{6}\Big[6(\lambda^2-\lambda+\frac{1}{2})\Big(2\tau_{\sigma}-\tau_\sigma^{unif}-
\frac{1}{36}\frac{(\vec{\nabla}\rho_{\sigma}(\mathbf{r}))^2}{\rho_{\sigma}(\mathbf{r})}\Big)\nonumber\\
+ \frac{6}{5}k_{\sigma}^2\rho_{\sigma}(\mathbf{r})(f_{\sigma}^2-1)-2\gamma D_{\sigma}\Big].\nonumber\\~.
\end{eqnarray}
Following BR approach \cite{br89}, the comparison the exchange hole with the exact exchange hole of the hydrogen atom 
(which is analytically derivable), we have arrived at the following one-dimensional nonlinear equation
\begin{equation}
\frac{x\exp(-2x/3)}{x-2} = \frac{2}{3}\pi^{2/3}\frac{\rho_{\sigma}^{5/3}}{Q_{\sigma}}.
\label{nonlinear}
\end{equation}
This nonlinear equation can be solved numerically by using efficient numerical root finding technique. For each value of 
density, the gradient of density and kinetic energy density we have found the positive root for $x$. As a matter of which, the 
coordinate transformed exchange potential is obtained as
\begin{eqnarray}
U^t_{x\sigma}(\mathbf{r}) &=& -(1-e^{-x}-\frac{1}{2}xe^{-x})/b
\label{pot1}
\end{eqnarray}
with,
\begin{equation}
b^3 = \frac{x^3\exp(-x)}{8\pi\rho_{\sigma}}.
\end{equation}
The unique feature of this exchange hole is that the underlying potential decays as $\approx -\frac{1}{r}$ when 
$r\to\infty$. Finally, the exchange energy is obtained from the parametrized exchange potential as,
\begin{equation}
E^t_{X\sigma} = \frac{1}{2}\int\rho_{\sigma}U^t_{x\sigma}(\mathbf{r})~d^3r.
\end{equation}

\section{Results and Discussion}
We consider Tran-Blaha modified Becke-Johnson potential (TBMBJ)~\cite{tbmbj} with our modified Becke-Roussel 
potential,
\begin{equation}
v_{\text{x}\sigma}^{\text{mBR@TBMBJ}}(\mathbf{r}) = cv_{\text{x}\sigma}^{\text{mBR}}(\mathbf{r}) +
\left(3c-2\right)\frac{1}{\pi}\sqrt{\frac{5}{12}} \sqrt{\frac{2\tau_\sigma(\mathbf{r})}
{\rho_\sigma(\mathbf{r})}},
\end{equation}
where $c$ is given by
\begin{equation}
c = -\mathcal{A}+\mathcal{B}\sqrt{\bar{\mathcal{G}}}
\end{equation}
and
\begin{equation}
\bar{\mathcal{G}}= \frac{1}{\mathcal{V}_\text{cell}}\int_{cell}\frac{1}{2}\Big[\sum_{\sigma}\frac{|\nabla
\rho_\sigma(\mathbf{r})|}{\rho_\sigma(\mathbf{r})}\Big]~d^3r
\end{equation}
is the average of spin-polarized $\frac{|\nabla\rho_\sigma(\mathbf{r})|}{\rho_\sigma(\mathbf{r})}$ over 
the unit cell of volume ${\mathcal{V}}_\text{cell}$. The modified Becke-Roussel potential is given by 
$v_{\text{x}\sigma}^{\text{mBR}}(\mathbf{r})=U^t_{x\sigma}(\mathbf{r})$. $\mathcal{A}$ and $\mathcal{B}$ 
are two parameters that have to  be fixed latter along with $\lambda$ and $\beta$. 

\subsection{Performance of modified Becke-Roussel for atomic systems}
\begin{table}
\caption{Exchange energies of noble-gas atoms (a.u.)}
\begin{tabular}{c  c  c  c  c  c  c  c} 
\hline\hline
Atoms&HF&LDA&BR\cite{br89}&BR\cite{br89}& mBR\\ 
     &     &   &$\gamma=1.0$&$\gamma=0.8$&$\gamma=1.0$\\ 
     &     &   &            &            &$\lambda=0.877$\\ 
     &     &   &            &            &$\beta=20.0$\\ \hline
He& -1.026&-0.884  &-1.039&-1.039 & -1.022  \\
Ne& -12.11&-11.03  &-12.19&-12.33 & -12.254   \\
Ar& -30.19&-27.86  &-30.09& -30.55& -30.353  \\
Kr& -93.89&-88.62  &-92.88&-94.77 & -93.898   \\
Xe& -179.2&-170.6  &-176.4&-180.3 &-178.681    \\
\hline
MAE($\Delta$)& &8.8 &1.1 &1.2 &{\bf{0.82}}\\
\hline\hline
\end{tabular}
\end{table}

\begin{figure}[!htbp]
\begin{center}
\includegraphics[width=3.2in,height=2.3in,angle=0.0]{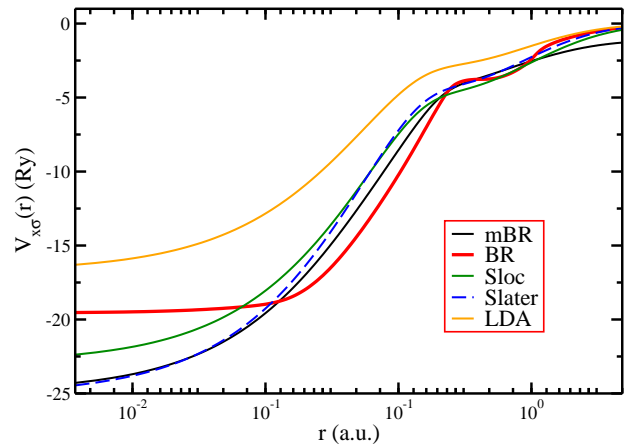} 
\end{center}
\caption{The exchange potential of Ne atom using LDA, BR and mBR formalism is compared with the Slater and local 
slater potential (Sloc) of it.}
\label{Ne-apot}
\end{figure}
We will first apply our proposed modified BR (mBR) model to calculate the exchange energy of noble gas from He 
through Xe. The parameter $\lambda$ is obtained by matching with the exact exchange for H atom (i.e., $0.312$ eV). 
Tao-Mo also followed the same strategy to fix $\lambda$ value \cite{tm16}. Slightly adjustable value of $\beta$ 
also, confirm the almost exact exchange energy for He atom. We obtained $\lambda=0.877$, $\gamma=1$ and $\beta=20.00$ 
in our present model. Atomic calculation for our functional is performed using deMon2k~\cite{demon} code with 
{\bf{DZVP}} basis set~\cite{basis}. We implemented an accurate Newton's algorithm~\cite{booknr} within deMon2k 
code for root finding of mBR potential. The value of exchange energy along with the mean absolute error (MAE) for 
the noble gas atoms is given in Table-I. Results show that the proposed functional performs much better than the 
original BR functional. The explanation of the improvement of exchange energy using the mBR over BR potential for 
atoms is simple. Atomic systems are always localized and by using the generalized coordinate transformation we 
make the exchange hole more localized. Therefore, we obtain improved exchange energy for mBR compared to the BR,
because of the inclusion of inhomogeneity through $k$ vector the mBR model perform more accurately than original 
BR.

\begin{figure}[!htbp]
\begin{center}
\includegraphics[width=3.2in,height=2.3in,angle=0.0]{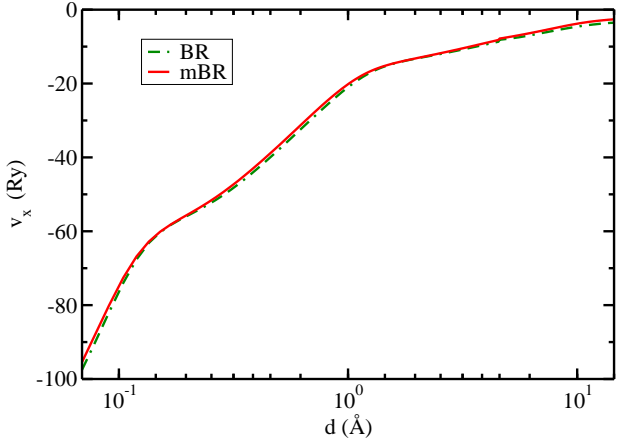} 
\end{center}
\caption{Shown are the exchange potentials $v_x$ for original BR and the newly constructed parametrized BR of Kr 
calculated using WIEN2k code.}
\label{kr-pot}
\end{figure}

In Fig.-(\ref{Ne-apot}) we compare our modified BR potential with that of original BR, LDA, Slater and local 
potential proposed (Sloc) by Kati Finzel et.al.~\cite{kf}. We named it Sloc as it was used by F. Tran et.al.
~\cite{ftjpca}. It consists of enhanced  LDA exchange fitted with the Slater exchange potential ($v_x(\mathbf{r})
=-1.67\rho(\mathbf{r})^{0.3}$). LDA is the local potential obtained from the functional derivative of exchange 
energy of homogeneous electron gas ($v_x^{LDA}(\mathbf{r})=-0.7386\rho(\mathbf{r})^{1/3}$). Here the Slater 
potential is defined as $v_x^{Slater}(\mathbf{r})=\frac{3}{2}v_x^{LDA}(\mathbf{r})$. The lower and middle 
portion of the mBR potential proposed by us matches well with that of Slater and Slocal potential compared to 
the BR potential. The step structure of the mBR potential also resembles well with Sloc and Slater potential. 
It has been observed that Slater potential has different nature of step structure than Exact exchange~\cite{HS1990}. 
Also, the nature of BR potential near nucleus is very different from that of other potentials. This is due to the 
fact that BR potential has Laplacian of density as one of its ingredient. Near the nucleus Laplacian of density 
diverges, which makes the root of nonlinear Eq.(\ref{nonlinear}) constant. This corresponds to the constant 
value of exchange potential from Eq.(\ref{pot1}). That's why we find the constant value of BR potential near the nucleus.
Therefore, BR potential failed to achieve the behavior of Slater or LDA potential near the nucleus. In that sense 
our Laplacian free modification over BR is quite appropriately produces the nature of Slater near the nucleus and 
at the core. Hence, our modified BR potential is a good semilocal substitution of Slater potential.

\subsection{Band Gaps}
The band gap calculations for different semiconductors, insulators, metal oxides are performed using WIEN2k 
code~\cite{wien2k}. We implemented our modified Becke-Roussel by locally modifying the Becke-Roussel part of 
TBMBJ functional implemented in WIEN2k code. In WIEN2k the root finding technique of BR functional is 
substituted by an analytic representation of BR functional, which produces same results as root finding 
technique. The WIEN2k code is most efficient and accurate among the available codes to calculate band structures. 
It uses Full Potential Linearized Augmented Plane Wave (FP-LAPW) method to calculate the ground state properties 
and electronic structure of solids. As no energy expression is available for this model potential, Tran-Blaha 
suggested that self-consistence TBMBJ calculation followed by a self-consistence PBE calculation. The space 
group and experimental geometries used for PBE calculations are given in Table-(\ref{band-table}). 

\begin{table*}
\label{band-table}
\caption{Systems, structures and experimental references of the solids considered in this work. Fundamental 
band gap (in eV) calculated at the experimental geometry. HSE06, G$_0$W$_0$ and GW results are taken from the 
literature are indicated. For TBMBJ and mBR@TBMBJ we have carried out self-consistent calculation using WIEN2k 
code. All TBMBJ and mBR@TBMBJ calculations are followed by self-consistence PBE calculations, given in $5^{th}$ 
column.}
\begin{tabular}{c  c  c  c  c  c  c  c  c  c  c} 
 \hline\hline
 Solid&structure&Space Group&Geometry(A)&PBE&TBMBJ&mBR@TBMBJ&HSE06$^b$&G$_0$W$_0$&GW&Expt.\\\hline\hline
Ar&fcc & $Fm\bar{3}m$& 5.310&8.676&14.288 &16.700 &10.37 &13.28$^a$ &14.9$^a$ &14.2\\
Kr&fcc & $Fm\bar{3}m$ & 5.650&7.259 &10.917  &12.607  &8.71 & & &11.6\\
Xe&fcc & $Fm\bar{3}m$  &6.130&6.260 &8.456  &9.749&7.44& & &9.8\\
C&Diamond & $Fd\bar{3}m$& 3.567&4.167 &4.966  &5.148 &5.26 &5.50$^a$ &6.18$^a$ &5.48\\
Si&Diamond &$Fd\bar{3}m$ & 5.430&0.581 & 1.162 & 1.302&1.17  &1.12$^a$ &1.41$^a$ &1.17\\
Ge&Diamond & $Fd\bar{3}m$ & 5.652&0.058 &0.824  &0.962  &0.82 &0.66$^a$ & 0.95$^a$&0.74\\
LiF&Rock salt & $Fm\bar{3}m$& 4.010&9.195 &13.035&13.895 &11.46  &13.27$^a$ &15.9$^a$ &14.2\\
LiCl&Rock salt &$Fm\bar{3}m$ & 5.106&6.366 &8.705  &9.698 &7.81  & & &9.4\\
MgO&Rock Salt & $Fm\bar{3}m$  &4.207&4.786 & 7.226& 7.492  &6.47 &7.25$^a$ &9.16$^a$ &7.83\\
BaSe&Rock Salt & $Fm\bar{3}m$&6.595&1.990 &2.893 &3.254   &2.79 & &&3.58\\
BaTe&Rock Salt & $Fm\bar{3}m$&7.007&1.604 &2.277 &2.558   &2.31 & &&3.08\\
BaS&Rock Salt &$Fm\bar{3}m$ & 6.389&2.218 & 3.326& 3.746& 3.11  && &3.88\\
MgS&Zinc blende & $F\bar{4}3m$&5.622&3.507 &5.16 &5.718 &4.66  & & &5.4  \\
SiC&Zinc blende & $F\bar{4}3m$&4.358&1.360 &2.278 &2.375&2.23  &2.27$^a$ &2.88$^a$ & 2.4  \\
BN&Zinc blende &  $F\bar{4}3m$&3.616&4.470 &5.816 &6.093&5.76  &6.10$^a$ &7.14$^a$ & ~6.25  \\
ZnS&Zinc blende & $F\bar{4}3m$&5.409&2.076 &3.692 & 3.765&3.30&3.29$^a$ &4.15$^a$ &3.91  \\
GaN&Zinc blende & $F\bar{4}3m$&4.523&1.658 & 2.7& 2.709&2.85  &2.80$^a$ &2.88$^a$ &3.2  \\
GaAs&Zinc blende & $F\bar{4}3m$&5.648&0.534 &1.886 &1.759&1.40  &1.30$^a$ &1.85$^a$ &1.52   \\
CdS&Zinc blende &  $F\bar{4}3m$&5.818&1.143 &2.601 &2.96&2.14  &2.06$^a$ &2.87$^a$ &2.42   \\
AlP&Zinc blende & $F\bar{4}3m$&5.463&1.587 &2.291 &2.617&2.30  &2.44$^a$ &2.90$^a$ &2.45   \\
AlAs&Zinc blende & $F\bar{4}3m$&5.661&1.445 &2.142 &2.37& 2.11 & & & 2.23  \\
BP&Zinc blende & $F\bar{4}3m$&4.538&1.246 &1.84 &1.958&1.98  & & & 2.4  \\
BAs&Zinc blende & $F\bar{4}3m$&4.777&1.182 &1.646 &1.813&1.86  & & & 1.46  \\
AlSb&Zinc blende & $F\bar{4}3m$&6.136&1.241 &1.781 &1.912&1.80  & & & 1.68  \\
GaSb&Zinc blende & $F\bar{4}3m$&6.096&0.113 &0.986 &1.014&0.88  & & &0.73   \\
InP&Zinc blende & $F\bar{4}3m$&5.869&0.675 &1.591 &1.833&1.43  & & &  1.42 \\
ZnSe&Zinc blende & $F\bar{4}3m$&5.668&1.263 & 2.71&2.911&2.37  & & & 2.7  \\
ZnTe&Zinc blende & $F\bar{4}3m$&6.089&1.251 & 2.402&2.501&2.25  & & & 2.38  \\
CdSe&Zinc blende & $F\bar{4}3m$&6.052&0.592 &1.907 & 2.228&1.52  & & & 1.9 \\
AlN&Wurtzite & $P63mc$&a=3.111, c=4.978&4.316 &5.750&5.748&5.49  &5.83$^a$ & &6.28   \\
ZnO&Wurtzite &  $P63mc$ & a=3.350, c=5.220&0.830 &2.65 & 2.493&2.50  &2.51$^a$ &3.8$^a$ & 3.44 \\
SiO$_2$&Rutile & $P42mnm$& a=4.181, c=2.662&5.861 &7.7 & 7.689&7.39  & & & 8.9 \\
SnO$_2$&Rutile & $P42mnm$ & a=4.740, c=3.190&1.569 &3.432 &3.419&2.88  & & &3.6   \\
TiO$_2$&Rutile &  $P42mnm$ &a=4.594, c=2.959&1.953 &2.667 &2.732&3.34   &3.34$^c$ &3.34$^c$ &3.3   \\
Cu$_2$O&Cuprite & $Pn\bar{3}m$ & a=4.267& 0.622 &0.934 &0.940&1.98   &1.97$^c$ &1.97$^c$ &2.17   \\
SrTiO$_3$&Perovskite& $Pm\bar{3}m$ & a=3.905&2.012&2.89 &2.921&3.29  &5.07$^c$ &5.07$^c$ &3.3   \\
\hline\hline
\end{tabular}
\footnotetext[1]{Reference \cite{tbmbj}}
\footnotetext[2]{Reference \cite{ftjpca}}
\footnotetext[3]{Reference \cite{KTB2011}}
\label{band-table}
\end{table*}

\begin{table}
\caption{Summary Statistics for the Error in the Calculated Band Gap for the Set of Solids listed in 
Table-(\ref{band-table}) }
\begin{tabular}{c  c  c  c  c  c  c  c  c  c  c}
\hline\hline
      &PBE&TBMBJ&mBR@TBMBJ&HSE06&G$_0$W$_0^{a}$&GW$^b$\\\hline\hline
ME(eV)&-1.80   &-0.36      &-0.02      &-0.69     &-0.23              &0.55      \\
MAE(eV)&1.80   &0.44      &0.44      &0.74     &0.44              &0.61      \\
STDE(eV)&1.24    &0.46      &0.64      &0.95     &0.60              &0.59      \\
MRE(\%)&-46.0   &-6.0      &1.0      &-10.0     &-5.0              &14.0      \\
MARE(\%)&46.0   &12.0      &14.0      &14.0     &11.0              &16.0      \\
STDRE(\%)&17.0   &15.0      &18.0      &13.0     &16.0              &15.0      \\ \hline\hline
\end{tabular}
\footnotetext[1]{18 data set}
\footnotetext[2]{17 data set}
\label{error}
\end{table}

\begin{figure}
    \centering
    \begin{minipage}{.5\textwidth}
        \centering
        \includegraphics[width=0.9\linewidth, height=0.3\textheight]{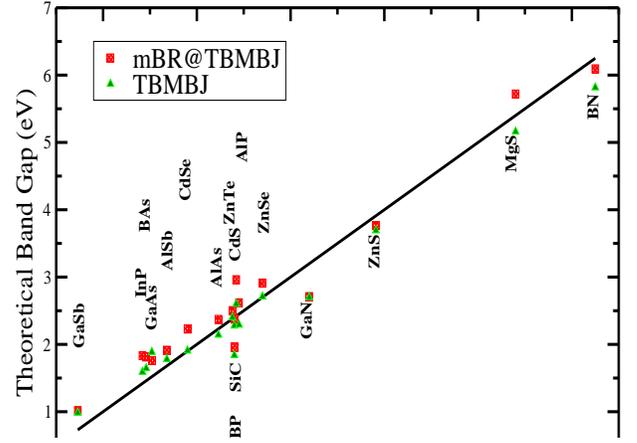}
        \centering
        \includegraphics[width=0.9\linewidth, height=0.3\textheight]{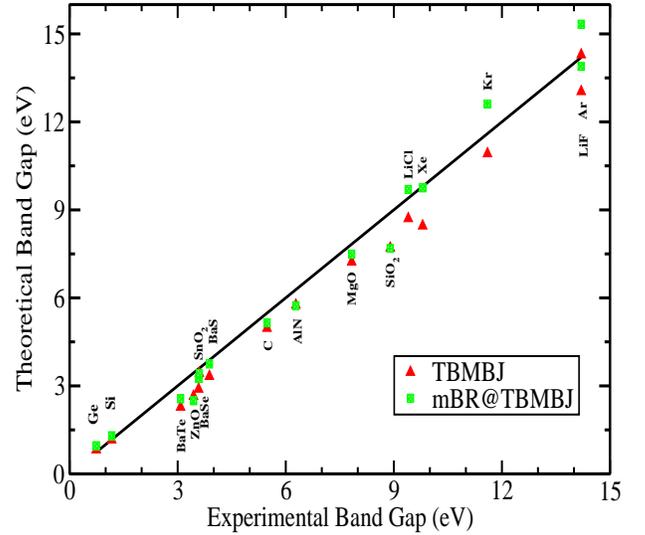}
        \caption{Theoretical versus experimental band gaps of all zinc blende structures 
        (upper panel) and all the structures except zinc blende structures (lower panel) 
        presented in Table- (III)}
\label{fig1}
    \end{minipage}
\end{figure}
\begin{figure}
\begin{center}
\includegraphics[width=2.5in,height=3.0in,angle=0.0]{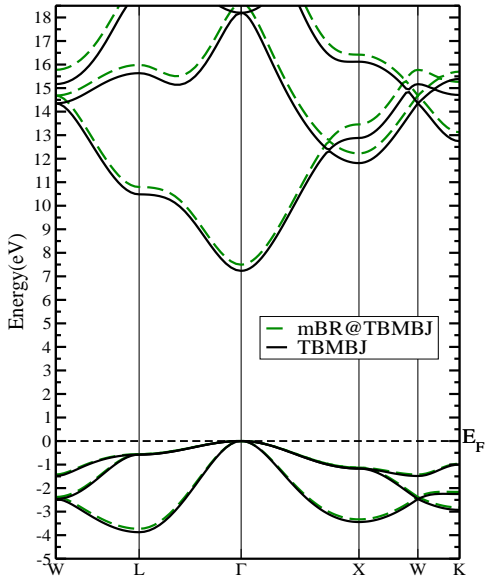} 
\end{center}
\caption{Band Structure of MgO obtained from TBMBJ (black solid) and our modified Becke-Roussel at TBMBJ 
(green dotted) calculations. The Fermi energy is at zero.}
\label{fig3}
\end{figure}

\begin{figure}
\begin{center}
\includegraphics[width=2.5in,height=3.0in,angle=0.0]{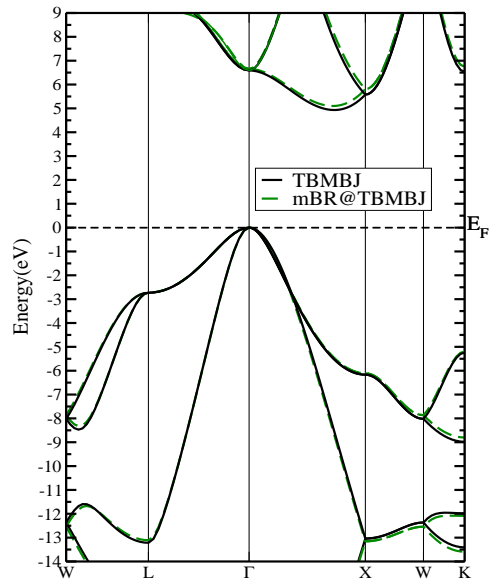} 
\end{center}
\caption{Band Structure of C obtained from TBMBJ (black solid) and our modified Becke-Roussel at TBMBJ 
(green dotted) calculations. The Fermi energy is at zero.}
\label{fig4}
\end{figure}

\begin{figure}
\begin{center}
\includegraphics[width=3.2in,height=2.3in,angle=0.0]{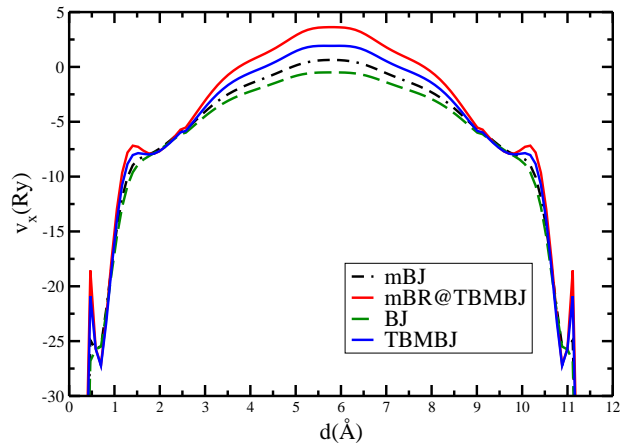} 
\end{center}
\caption{Exchange potentials $v_x$ in Xe plotted for different semilocal potentials.}
\label{fig5}
\end{figure}

\begin{figure}
\begin{center}
\includegraphics[width=3.2in,height=2.3in,angle=0.0]{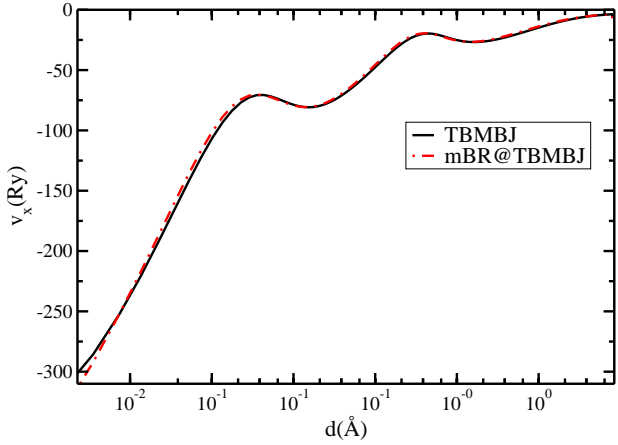} 
\end{center}
\caption{Exchange potentials ($v_x$) plotted for Cu$_2$O from Cu atom at site (1/2,1/2,0) towards O atom at 
(3/4,3/4,3/4). For $x$-axis we used logarithmic scale.}
\label{fig6}
\end{figure}

\begin{figure}
\begin{center}
\includegraphics[width=3.5in,height=3.0in,angle=0.0]{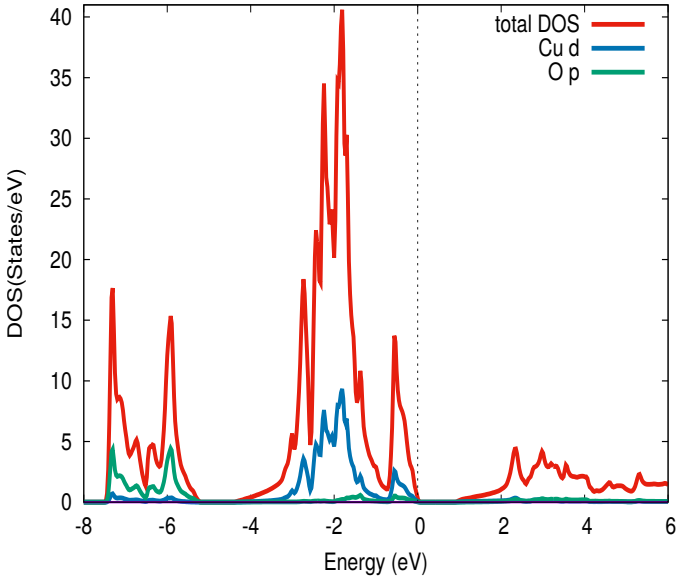} 
\end{center}
\caption{DOS and density of states for individual orbitals for Cu$_2$O for mBR@TBMBJ potential. Fermi energy 
is set at zero.}
\label{fig-dos}
\end{figure}

\begin{figure}
\begin{center}
\includegraphics[width=3.2in,height=2.3in,angle=0.0]{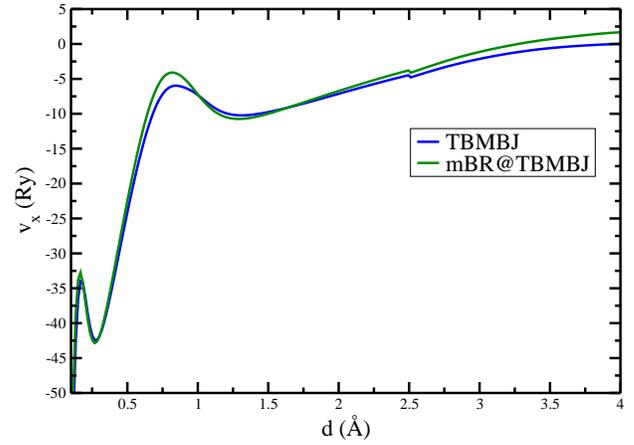} 
\end{center}
\caption{Exchange potentials ($v_x$) plotted for LiCl from Li  towards Cl.}
\label{licl-pot}
\end{figure}

\begin{figure}
\includegraphics[width=.45\linewidth]{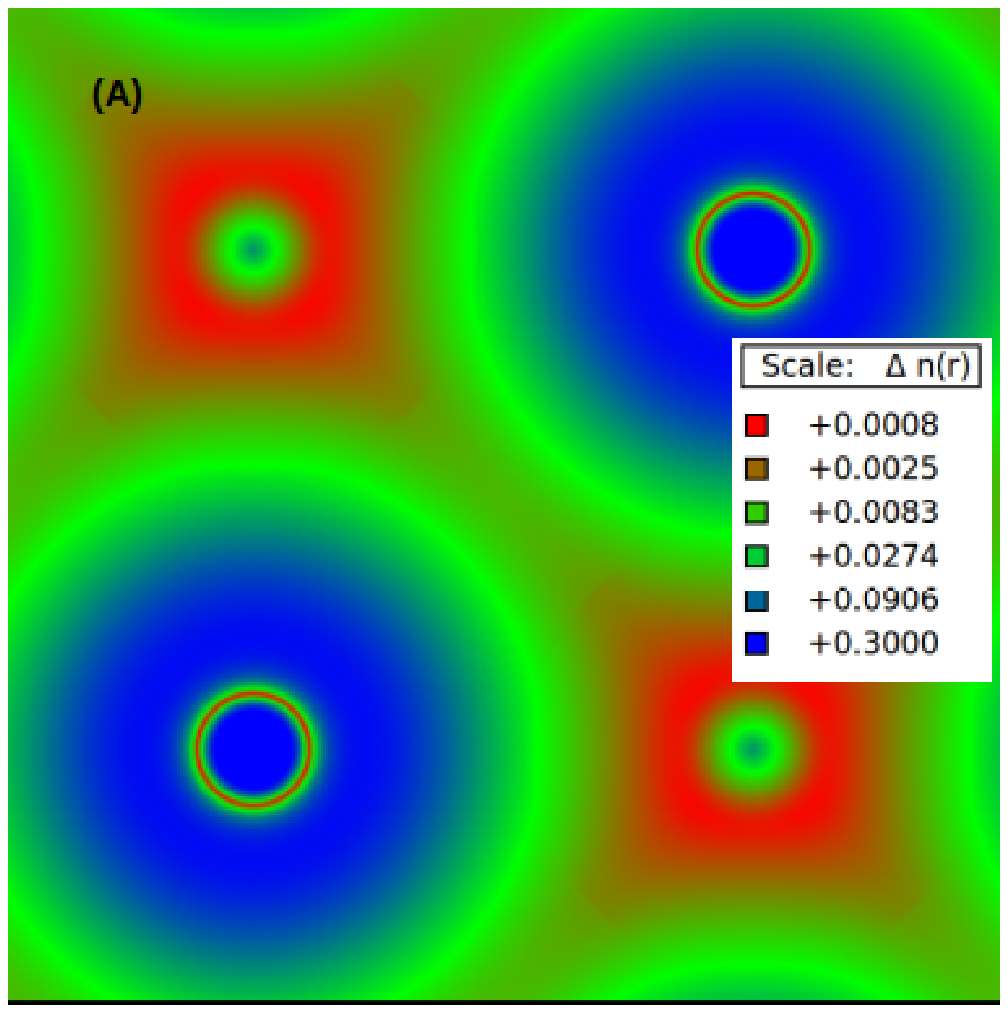}\hfill
\includegraphics[width=.45\linewidth]{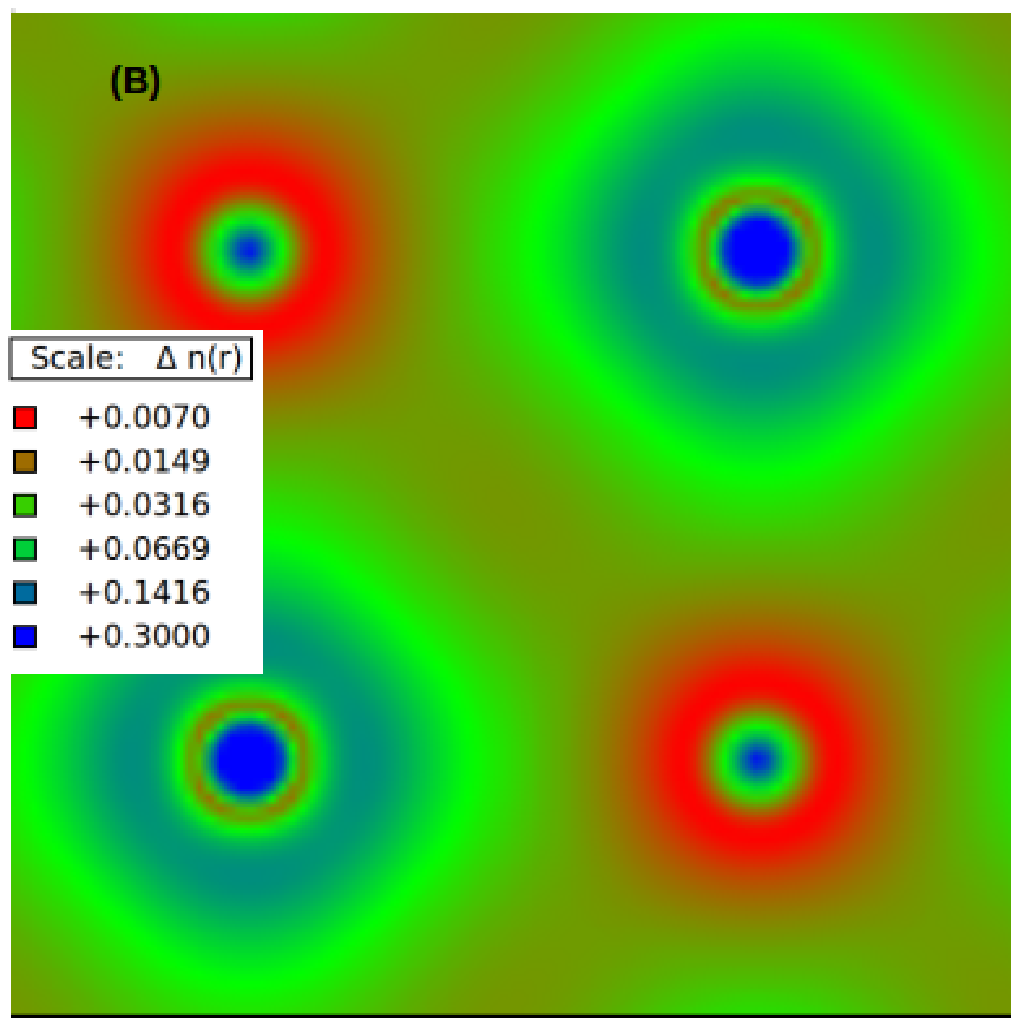}
\includegraphics[width=.45\linewidth]{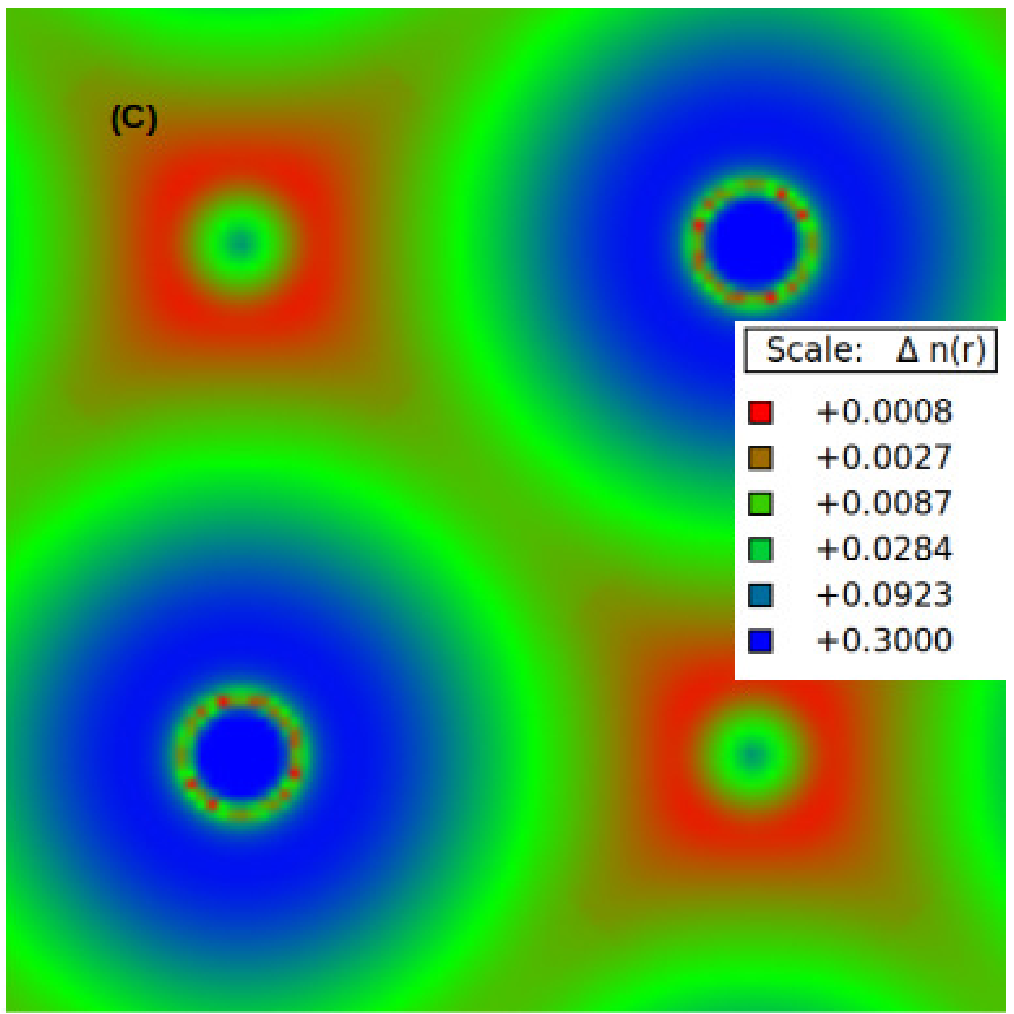}\hfill
\includegraphics[width=.45\linewidth]{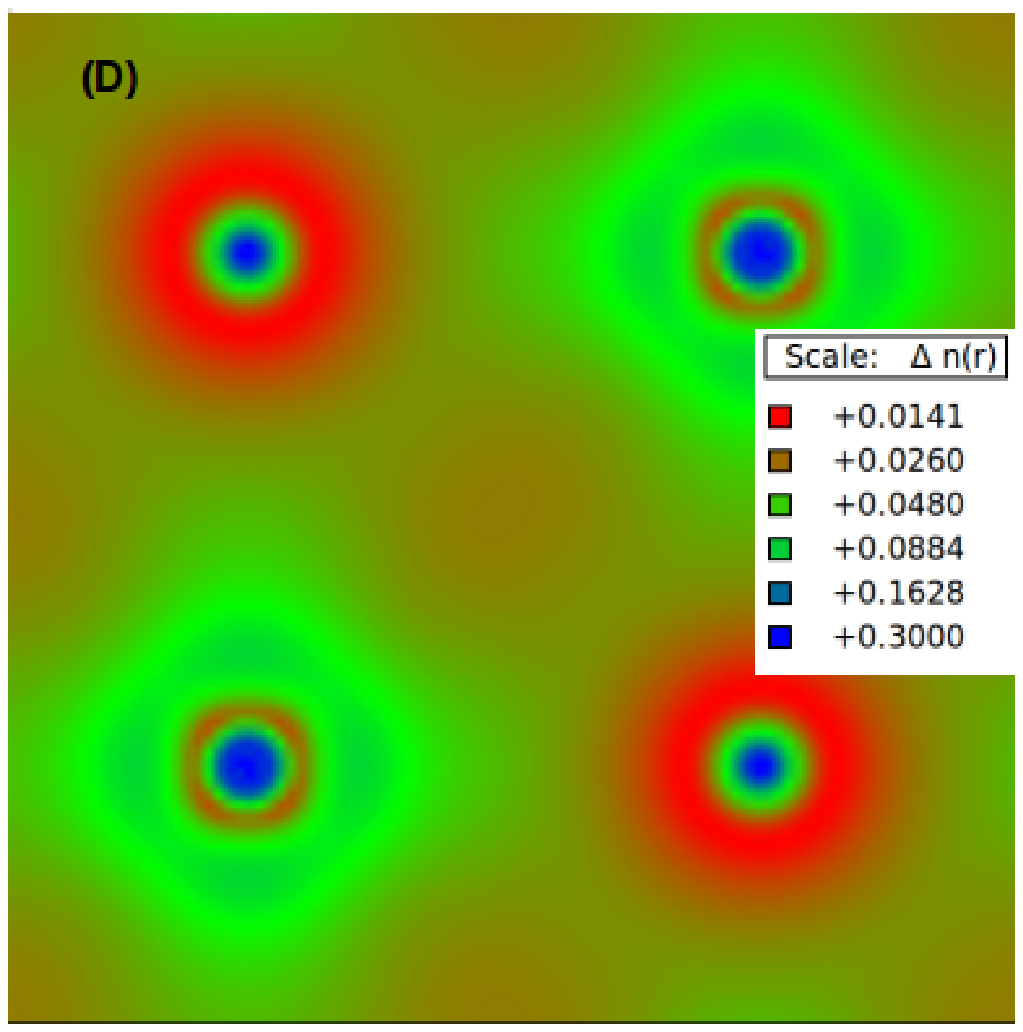}
\caption{Electron density (in e/bohr$^3$) of LiCl  (A) VB of mBR@TBMBJ, (B) CB of mBR@TBMBJ (C) VB of TBMBJ, 
(D) CB of TBMBJ. logarithmic scale used for this plotting.}
\label{lich-ch-den}
\end{figure}

The constants $\mathcal{A}$ and $\mathcal{B}$ present in the mBR@TBMBJ potential is fixed by matching 
with the experimental band gaps. We obtain optimized value $\mathcal{A}=0.030$ and $\mathcal{B}=1.0$. 
Whereas, the original TBMBJ used the optimized value, $\mathcal{A}=0.012$ and $\mathcal{B}=1.023$. 
Several other modifications over  original TBMBJ  also proposed by Koller {\it{et.al.}} \cite{KTB2011}. 
We compared our band gap results with that of original TBMBJ potential. The performance of hybrid 
HSE06~\cite{pbe0,jcphse}, G$_0$W$_0$ and GW is also shown in Table-(\ref{band-table}). A plot of all the 
theoretical versus experimental band gap for zinc blende and other structures are shown separately in 
Fig.-(\ref{fig1}). Our results for the semiconductors, insulators, transition-metal oxides and other 
transition-metal oxides that are considered here and previously reported by other researchers indicate 
that there is a significant reduction in band gap error compared to the original TBMBJ potential. 
For Ar, Kr our functional overestimated band gap almost by $1$ eV but for Xe we obtained results very 
close to the experimental value. Among diamond structures slightly overestimation of band gap is observed 
for Si and Ge but for C we obtained good band gap of $\approx$ $5.15$ eV compared to its experimental 
value of $5.48$ eV. The other parametrization proposed by Koller et.al. \cite{KTB2011} produced the highest 
band gap up to $5.00$ eV. Thus, our parametrized potential improved the band gap of C w.r.t other existing 
parametrizations. The performance of our modification for obtaining the band gap of LiF, LiCl, MgO, BaSe, 
BaTe, BaS show that our proposed functional have outperformed TBMBJ and HSE functional. For SiC, BN, ZnS, 
BP, our results are closer to the diagonal line as shown in Fig.(\ref{fig1}). We obtain band gap close to 
experimental value for those solids. Whereas, for GaN, GaAs, GaSb we observed almost same gap as original 
TBMBJ. Among other zinc blende structures, our modification slightly overestimated gap with maximum 
overestimation of $0.5$ eV for Cds. The mean error (ME) for zinc blende band gap is still better than TBMBJ 
for our potential (shown in Table-(\ref{error})). Except for very few cases, the band gap is slightly 
overestimated, rather than the underestimation did by TBMBJ. The new functional have produced an almost same 
result as original TBMBJ for other structures. The only exception is observed for Wurtzite ZnO. In this 
case, our functional has estimated the band gap same as obtained by HSE06 and $G_0W_0$. The band structure 
plots for MgO and C along with the high symmetry point is given in Fig.(\ref{fig3}) and Fig(\ref{fig4}).
This clearly indicates that there is a rigid displacement of the conduction band along the higher energy 
level. Therefore, increase in band gap compared to TBMBJ is observed.

For further physical insight of the proposed exchange hole or potential, we compared the mBR and BR exchange 
potentials of Kr in Fig.-(\ref{kr-pot}). It shows that mBR potential matches well with BR in this case. Also 
the exchange potentials for Xe and Cu$_2$O are shown in Fig-(\ref{fig5}) and Fig-(\ref{fig6}) respectively. 
In Fig.-(\ref{fig5}) we have plotted the mBR with Becke-Johnson correction (mBR@TBMBJ), modified BR potential 
with MBJ correction and all the concerned unmodified versions. It is indicative that mBR@TBMBJ model pushes 
the potential up recognized as interstitial region and therefore, results to increase in the band gap w.r.t.
TBMBJ. The increase in potential is coming from mBR. In the interstitial region the mBR@TBMBJ is more repulsive 
than TBMBJ. Thus, results to shifting in the conduction band and increase in band gap. The shell structure of 
the new potential is more evident than TBMBJ and it can be understood through quantum mechanical interpretation 
as given by Harbola-Sahni~\cite{HS1989}. The explanation of increase in band gap due to upward shifting of the 
potential was also explained by F. Tran et.al. ~\cite{ftjctc,ftprb}. In Fig.-(\ref{fig6}), we have shown the 
comparison of our functional with TBMBJ functional. In this case our functional matches perfectly with that of 
TBMBJ. Thus our Laplacian free functional mimic the TBMBJ functional very well in this case. In~\cite{ftprb},
they have proposed different parametrization over BJ potential was known as generalized BJ which can also be tested 
with our modified functional. In Figures-(\ref{fig-dos}) the DOS of Cu$_2$O is shown. The partial DOS for Cu-$3$d 
is obtained from our functional dominates within the range from $-3$ eV to $0$ eV below the Fermi level. Also 
O-$2$p DOS extended from $-7$ eV to $-5$ eV. This is an inherent difficulty for TBMBJ to estimate the band gap of 
Cu$_2$O accurately. This problem also remains unresolved by our modification. As the addition of BJ correction 
underestimates the band gap of Cu$_2$O, therefore our modification over BR potential unable to increase the band 
gap significantly. A test for band gap using only mBR potential may be performed. Thus we argue that why our 
functional produced better band gap than TBMBJ in some case and why it is producing similar results to that of 
TBMBJ (for example Cu$_2$O). Our comparison of exchange potentials clearly indicates that the new construction 
provides a noteworthy improvement over TBMBJ in most cases. 

Next, for an example, we prefer rock salt LiCl structure to encapsulate the changes in the electron density 
influenced by TBMBJ and our modified BR with TBMBJ exchange potential (shown in Fig.(\ref{lich-ch-den})). 
The physical explanation of opening of band gap for LiCl, which is an ionic compound can be explained by 
the nature of the exchange potential is shown in Fig.(\ref{licl-pot}). The nature of the potential shows that 
in valence region both the potential behaves in the same manner, but in the conduction band states (recognized 
as the interstitial region) there is an upward shift of the mBR@TBMBJ potential. Due to the different nature of 
both the potentials, there are changes in the conduction band density distribution for mBR@TBMBJ compared with 
TBMBJ exchange potential. In valance band,  there is no difference in density distribution for both the potential 
as shown in Fig.(\ref{lich-ch-den}). In the conduction band states, due to the upward shift of the mBR@TBMBJ 
potential electron density around Cl$^-$ ion enhanced, therefore enhancing the ionic character around Cl$^-$ 
by increasing the number of electrons around Cl$^-$. The nodal plane distribution of density for both the 
exchange potential stipulates that for valance band the density distribution for both the exchange potentials 
are same but for conduction band the nodal structure of mBR@TBMBJ potential more enhanced around Cl$^-$ compared 
to TBMBJ potential.

\section{Conclusion}
To summarize, we have proposed a modified version of BR potential by substituting gradient expansion of Kohn-
Sham kinetic energy in place of the Laplacian of density present in the exchange hole. Also included the extra 
inhomogeneity present in the system through the momentum vector without hindering the uniform density limit. 
The proposed functional is also parametrized by making use of the generalized coordinate transformation. 
Therefore, the newly constructed model potential is placed once step ahead of the BR potential. The parameter 
present in the modified BR potential is fitted with the atomic systems. The proposed functional has achieved 
better accuracy over original BR functional. Next, we have studied the band gaps of solids by combining the  
proposed exchange hole potential with that of the TBMBJ. Our demonstration shows that there is a significant 
improvement in the calculation of band gaps. This shows why the present modification over the BR potential 
could be a good approximation for Slater potential. Next, step of this work is to investigate the $\sqrt{
{\tau}/\rho}$ dependency of the new functional. There are many modifications/developments which have been 
proposed over Becke-Johnson potential~\cite{kummel1,kummel2,gs1,gs2,gs3,rasa1,rasa2,rasa3}. But, our present 
attempt is to present a more realistic and accurate exchange hole model by keeping intact the associated 
formal properties. In solid state calculations, it is also necessary to recover the slowly varying LDA limit. 
The intricate balance between the recover of LDA limit and localization can be produced by localized exchange 
hole which is capable of obtaining approximately LDA exchange potential. Also the localized exchange 
hole make the HOMO and LUMO localized, therefore, rectify the underestimation of band gap slightly. Thus it is 
always interesting to design a localized exchange hole (or potential). At the end, we want to conclude that 
besides the derivative discontinuity, the other inherent problems of the semilocal formalism is that they treat 
the electrons over delocalized due to the presence of self-interaction error, which makes the exchange hole also 
delocalized. Therefore, more insight of the effect of semilocal potential design from localized exchange hole 
and studying band gaps is our future plan of investigation.

\section{Acknowledment}
The financial support from the Department of Atomic Energy, Government of India is acknowledeged.

\end{document}